\documentclass[11pt]{article}
\usepackage{graphicx}
\usepackage{amssymb}
\usepackage{amsmath}
\usepackage{lscape}
\usepackage{fancyhdr}
\usepackage{floatrow}
\usepackage{natbib}
\usepackage{color, soul}
\usepackage{enumitem}
\soulregister\citet7
\soulregister\citep7

\begin{document}
\date{}


\begin{center}
\par\underline{\Large Voyage 2050 white paper}
\bigskip
\par\textbf{\huge The local dark sector}
\bigskip
\par\textbf{\Large Probing gravitation's low-acceleration frontier and dark matter in the Solar System neighborhood}
\bigskip
\bigskip


Joel Berg\'e$^1$, Laura Baudis$^2$, Philippe Brax$^3$, Sheng-Wey Chiow$^4$, Bruno Christophe$^1$, Olivier Dor\'e$^4$, Pierre Fayet$^5$, Aur\'elien Hees$^6$, Philippe Jetzer$^2$, Claus L\"ammerzahl$^7$, Meike List$^7$, Gilles M\'etris$^{8}$, Martin Pernot-Borr\`as$^{1,9}$, Justin Read$^{10}$, Serge Reynaud$^{11}$, Jason Rhodes$^4$, Benny Rievers$^7$, Manuel Rodrigues$^1$, Timothy Sumner$^{12}$, Jean-Philippe Uzan$^{9}$, Nan Yu$^4$


\bigskip
\bigskip


\begin{itemize}
\setlength\itemsep{0.07em}
\item[] $^1$ DPHY, ONERA, Paris Saclay University, 29 av. Division Leclerc F-92322 Ch\^atillon Cedex, France
\item[] $^2$ Physik-Institut, Universitaet Zuerich, Winterthurerstrasse 190, CH-8057 Zuerich, Switzerland
\item[] $^3$ Institut de Physique Th\'eorique, Universit\'e Paris-Saclay, CEA, CNRS, F-91191 Gif-sur-Yvette Cedex, France
\item[] $^4$ Jet Propulsion Laboratory, California Institute of Technology, 4800 Oak Grove Drive, Pasadena, CA, 91109, USA
\item[] $^5$ LPENS, Ecole Normale Sup\'erieure (Universit\'e PSL, CNRS, Sorbonne Universit\'e, Universit\'e de Paris, Paris, France), and CPhT, Ecole Polytechnique (Palaiseau, France)
\item[] $^6$ SYRTE, Observatoire de Paris, Universit\'e PSL, CNRS, Sorbonne Universit\'e, LNE, 61 avenue de l'Observatoire, 75014 Paris, France
\item[] $^7$ ZARM, Center of Applied Space Technology and Microgravity, University of Bremen, Am Fallturm, D-28359 Bremen, Germany
\item[] $^{8}$ Universit\'e C\^ote d{'}Azur, Observatoire de la C\^ote d'Azur, CNRS, IRD, G\'eoazur, 250 avenue Albert Einstein, F-06560 Valbonne, France
\item[] $^{9}$ Institut d'Astrophysique de Paris, CNRS UMR 7095, Universit\'e Pierre \& Marie Curie - Paris VI, 98 bis Bd Arago, 75014 Paris, France
\item[] $^{10}$ Department of Physics, University of Surrey, Guildford, GU2 7XH, Surrey, UK
\item[] $^{11}$ Laboratoire Kastler Brossel, UPMC-Sorbonne Universit\'e, CNRS, ENS-PSL University, Coll\`ege de France, 75252 Paris, France
\item[] $^{12}$ Blackett Laboratory, Imperial College London, Prince Consort Road, London. SW7 2AZ, United Kingdom
\end{itemize}

\end{center}

%
%
%
%
%
%

\bigskip
\bigskip
\bigskip

\newpage

\pagestyle{fancy}
\lhead{The local dark sector}
\rhead{}

\begin{center}
\par\textbf{\huge Executive summary}
\end{center}

We speculate on the development and availability of new innovative propulsion techniques in the 2040s, that will allow us to fly a spacecraft outside the Solar System (at 150 AU and more) in a reasonable amount of time, in order to directly probe our (gravitational) Solar System neighborhood and answer pressing questions regarding the dark sector (dark energy and dark matter). We identify two closely related main science goals, as well as secondary objectives that could be fulfilled by a mission dedicated to probing the local dark sector:

\paragraph{Science goal \#1}
{\it Begin the exploration of gravitation's low-acceleration regime with a man-made spacecraft.}  
When placed on a potential--curvature plane, experimental and observational probes of gravitation make apparent the existence of a ``desert" between the relatively high Solar System potential (where General Relativity --GR-- is well tested) and cosmological scale (where GR must be either replaced or complemented by dark energy). We have no clue of gravity's behavior in this intermediate regime.
Only by entering this potential--curvature desert with a dedicated spacecraft will we be able to gain new experimental clues about gravity, and perhaps about physics beyond GR. We could in particular enter a new intermediate regime that bridges between GR in the Solar System and ``dark energy gravitation"  at cosmological scales.

\paragraph{Science goal \#2}
{\it Improve our knowledge of the local dark matter and baryon densities}
The constraints that direct detection experiments can set on dark matter characteristics depend on the local dark matter density, which is degenerate with the dark matter particle's cross section. The estimation of this density in the Solar System neighborhood is performed by monitoring the motion of nearby stars, relying on numerical simulations and assuming that the dark matter distribution is homogeneous. However, the coupling between dark matter and baryons remains a significant caveat. Flying a spacecraft in the outskirts of the Solar System will allow for a direct measurement of the gravitational environment and will bring new and necessary information about the very local dark matter distribution and density, significantly reducing the need for the enormous extrapolation usually used from numerical simulations.

\paragraph{Science goals \#3}
{\it Additional goals allowed by a mission designed for Science goals \#1 and \#2}
We can define secondary science objectives about gravitation and fundamental physics (constrain post-Newtonian parameters and the evolution of fundamental constants), the Solar System (explore the Kuiper belt and the heliosheath) and gravitational waves (measure the gravitational wave stochastic background).

\bigskip
Those questions can be answered by directly measuring the gravitational potential with an atomic clock on-board a spacecraft on an outbound Solar System orbit, and by comparing the spacecraft's trajectory with that predicted by GR through the combination of ranging data and the in-situ measurement (and correction) of non-gravitational accelerations with an on-board accelerometer.

Despite a wealth of new experiments getting online in the near future, that will bring new knowledge about the dark sector, it is very unlikely that those science questions will be closed in the next two decades. More importantly, it is likely that it will be even more urgent than currently to answer them. Tracking a spacecraft carrying a clock and an accelerometer as it leaves the Solar System may well be the easiest and fastest way to directly probe our dark environment.

\newpage
\section{When the Universe goes dark}

General Relativity (GR) may arguably be the most elegant physics theory. It describes gravitation as the simple manifestation of spacetime's geometry, while recovering Newton's description of gravitation as a classical inverse-square law (ISL) force in weak gravitational fields and for velocities small compared to the speed of light. In its century of existence, GR has been extremely successful at describing gravity. From Mercury's perihelion to gravitational lensing, to gravitational waves direct detection, it has successfully passed all experimental tests (see e.g. \citet{will14} and \cite{ishak19} and references therein).
Most importantly, GR allowed for the advent of mathematical cosmology and is the cornerstone of the current hot Big Bang description of our Universe's history.

Sitting next to GR, the Standard Model (SM) was built from the realization that the microscopic world is intrinsically quantum. Although perhaps less elegant than GR because of its numerous free parameters, it is both highly predictive and efficient not only at describing the behavior of microscopic particles, but also at mastering key technologies. Increasingly large particle accelerators and detectors have allowed for the discovery of all particles predicted by the model, up to the celebrated Brout-Englert-Higgs boson \citep{aad12, chatrchyan12}.

Although both frameworks (GR and SM) leave few doubts about their validity and seem unassailable in their respective regimes (low vs high energy phenomena, macroscopic vs microscopic worlds), difficulties have been lurking for decades. Firstly, the question of whether GR and the SM should and could be unified remains open. Major theoretical endeavors delivered models such as string theory, but still fail to provide a coherent, unified vision of our world. Secondly, unexpected components turn out to make up most of the Universe's mass-energy budget: dark matter and dark energy are undoubtedly the largest conundrums of modern fundamental physics. 

\subsection{Dark matter} \label{sect_dm}
From observations of the Coma galaxy cluster, \cite{zwicky33} was the first to pinpoint the problem of missing matter: the galaxies motion could not be explained from the luminous matter only. At the onset of the 1980s, \cite{rubin80} noticed a similar behavior on galactic scales: galaxies rotate faster than expected based on their observed luminosity. Dark matter interacts only weakly with baryonic matter, but its gravitational interaction is necessary not only to account for galaxies' rotation curves and cluster galaxies dynamics, but also to explain peaks in the Cosmic Microwave Background (CMB) spectrum and the whole structure formation; no model is able to build galaxies without its presence.

Astronomers failed to find this missing matter as baryonic massive compact baryonic objects (MACHOs), making a stronger case for non-baryonic matter. Maps of the dark matter distribution at large scale, pioneered with weak gravitational lensing by \citet{massey07}, are now common (see e.g. \cite{vanwaerbeke13} or \citet{chang18}), and a cosmic web structured by (cold) dark matter is now central to the cosmological model.

Meanwhile, new approaches to gravitation, like MOND \citep{milgrom83}, were developed to explain observations without the need for dark matter. However, observations of colliding galaxy clusters \citep{clowe06} and of the fact that dark matter seems to be heated up by star formation in dwarf galaxies \citep{read19} tend to prove the existence of a particle non-baryonic dark component, albeit sterile neutrinos could also act as dark matter \citep{boyarsky19}.

While astronomers were looking for dark matter gravitational signatures, particle physicists did not remain idle. The supersymmetric extension of the SM predicts new massive particles, the lightest of which was expected to account almost exactly for the amount of dark matter in the Universe. Alas, this ``WIMP miracle" was a lure and has somewhat faded by now, after intensive indirect and direct searches failed to find it. Better and better direct detection experiments have been running for decades (see e.g. \citet{baudis16,schumann19} for recent reviews). Aside DAMA/LIBRA's observation of an annually modulated event rate, no experiment ever saw the signature of a WIMP hitting its detector. The axion, hypothesized as a by-product of a possible solution to the strong CP-problem in the 1970s \citep{kuster08}, is another good candidate still actively looked for, as are newer axion-like particles (ALPs). Nevertheless, only upper bounds on the WIMPs' mass and cross-section and on the axions and ALPs bounds to photon or electrons have so far been reported.

The current consensus remains the existence of a non-baryonic matter interacting with baryonic matter mostly gravitationally, that pervades the Universe and clumps as halos hosting galaxies and clusters of galaxies \citep{planck}. 

\subsection{Cosmological constant, dark energy and modified gravity} \label{sect_de}

The observation of the accelerated expansion of the Universe from the measurement of the distance of supernovae \citep{riess98, perlmutter99} added a new twist to cosmology and fundamental physics. Baryonic and dark matter were not alone, but a new component of the Universe was even dominating the mass-energy budget of the Universe. Since then, all cosmological probes have converged and their concordance allowed cosmologists to robustly infer the acceleration of the cosmological expansion \citep{planck, des18}.

The most natural candidate to explain this acceleration is Einstein's cosmological constant $\Lambda$ \citep{carroll01}, which may be linked to the vacuum energy. Adding it to cold dark matter, the $\Lambda$CDM cosmological model was born. However, the observed value of $\Lambda$ is around 120 orders of magnitude smaller than the naive expectation from Quantum Field Theory (QFT), that should be of the Planck mass: this {\it cosmological constant problem} is still unexplained.

The acceleration of the expansion can also be caused by a dynamical, repulsive fluid. Measuring the equation of state of this dark energy has become the grail of observational cosmology, with new and near future surveys: for instance (the list is far from exhaustive), the Dark Energy Survey has already published first results \citep{des18}, while ESA's Euclid\footnote{http://sci.esa.int/euclid/} and NASA's WFIRST\footnote{http://wfirst.gsfc.nasa.gov} missions are planned to finely measure weak lensing on half the sky, and provide percent constraints on dark energy within ten years.

In the dark energy view, GR keeps its central role as the theory of gravitation, assumed valid on all scales; it is the content of the Universe which is modified. The accelerated expansion can also be explained the other way around: no new component is added to the Universe, but GR is subsumed by a more general theory of gravitation that passes Solar System and laboratory tests while having a different behavior on cosmological scales. A plethora of models have been proposed (e.g. \citet{jain10, joyce15}), of which the scalar-tensor theories are the simplest.

GR describes the gravitational force as mediated by a single rank-2 tensor field. There is good reason to couple matter fields to gravity in this way, but there is no good reason to think that the field equation of gravity should not contain other fields. It is then possible to speculate on the existence of other such fields. The simplest way to go beyond GR and modify gravity is then to add an extra scalar field: such scalar-tensor theories are well established and studied theories of Modified Gravity \citep{damour92}. From a phenomenological point of view, scalar-tensor theories link the cosmic acceleration to a deviation from GR on large scales. They can therefore be seen as candidates to explain the accelerated rate of expansion without the need to consider dark energy as a physical component. Furthermore, they arise naturally as the dimensionally reduced effective theories of higher dimensional theories, such as string theory; hence, testing them can allow us to shed light on the low-energy limit of quantum gravity theories.

Scalar fields that mediate a long range force able to affect the Universe's dynamics should also significantly modify gravity in the Solar System, in such a way that GR should not have passed any experimental test. Screening mechanisms have been proposed to alleviate this difficulty \citep{joyce15}. In these scenarios, (modified) gravity is environment-dependent, in such a way that gravity is modified at large scales (low density) but is consistent with the current constraints on GR at small scale (high density). Several such models have been put forth, like the chameleon \citep{khoury04a, khoury04b}, the symmetron \citep{hinterbichler10}, the Vainshtein mechanism \citep{vainshtein72}, the K-Mouflage \citep{barreira15, brax15} or the dilaton \citep{brax10, damour10a, damour10b}. 

Those modified gravity models all predict the existence of a new, fifth force, that should be detectable through a violation of the ISL or of the equivalence principle. Despite numerous efforts, no such deviation has been detected so far (see e.g. \citet{fischbach99, adelberger03, touboul17, berge18} and references therein).

\subsection{Aim and layout of this white paper; mission concept summary}

Theoretical, observational and experimental efforts are still underway to shed light on dark energy or modified gravity. This white paper proposes new ways to peek more deeply than ever into this dark sector by directly probing it in the Solar System neighborhood.

To this aim, we propose to probe gravitation in the uncharted low-acceleration regime prevalent in the Solar System outskirts. An atomic clock onboard a spacecraft drifting out of the Solar System allows for a direct measurement of the local metric. Meanwhile, an onboard accelerometer measures all non-gravitational forces acting on the spacecraft, so that they can be substracted off and a ranging technique allows for a direct assessment of the geodetic motion of the spacecraft. Measuring the geodetic motion is equivalent to measuring the Levi-Civita connection (by comparison to clocks that directly measure the metric). Not only this mission concept allows for a test of gravity, but it also allows for the measurement of the local dark matter density disentangled from the baryonic density (as it creates measurable friction on the spacecraft), thereby helping to break the degeneracy between the local dark matter density and the dark matter cross section in direct detection experiments.

As the time required to fly a spacecraft farther than the Kuiper belt with traditional propulsion and space navigation techniques is prohibitively long, we will rely on new types of propulsion. We propose to use the concept elaborated for the Breakthrough Starshot project, using a solar sail illuminated by a high-power laser from the ground. Rough orders of magnitude allow us to expect to be able to send a probe of mass less than 1000 kg to hundreds of AU in two to four years, making it a viable mission. Recently, \citet{banik19} showed how a similar mission (though in a slightly different context) can be feasible.

Such a mission is expected to be a M-class mission, although it highly depends on the availability and cost to use the Breakthrough Starshot laser in the 2040s.

Sect. \ref{sect_sciencecase} presents, in as much details as possible, state-of-the-art experimental efforts to shed light on the aforementioned conundrums. We detail our science case in this section. Sect. \ref{sect_missionconcept} gives a broad outline of how a spacecraft (embarking an atomic clock, an accelerometer and tracked with a ranging technique) could perform it, and provides secondary science objectives that could easily be reached.
In Sect. \ref{sect_landscape}, we attempt to give a realistic landscape of the related science in the 2040s. We mention the expected technical challenges in Sect. \ref{sect_challenges}, then conclude in Sect. \ref{sect_conclusion}.

\section{Shedding light on the local Galactic dark sector} \label{sect_sciencecase}

In this section, we show how direct measurements of non-gravitational forces in the Solar System neighborhood (meaning, {\it outside} the Solar System) is a necessary step to go beyond current experimental constraints on gravitation, dark energy and dark matter.

\subsection{Dark energy and gravitation}

\subsubsection{Fifth force searches in the Solar System: current status}

All theoretical attempts at explaining the three limitations mentioned in Sect. \ref{sect_de} modify GR and Newton dynamics, either at small scale or very large scale, or both. In particular, if one of them is correct, we should detect a violation of the gravitational ISL. Hence, in the weak field limit, measurements of the dynamics of gravitationally bound objects (e.g. the behavior of a torsion balance in the Earth gravity field, the trajectory of an interplanetary probe, or the receding of galaxies) should show a deviation from what is expected from Newton's equations. In other words, a fifth force should be detectable \citep{fischbach99}.



It is common-practice to parametrize a deviation from the ISL with a Yukawa potential, such that the gravitational potential created by a point-mass of mass $M$ at a distance $r$ is \citep{fischbach99, adelberger03}
\begin{equation}
V(r) = -\frac{GM}{r} \left(1 + \alpha e^{-r/\lambda} \right),
\end{equation}
where $G$ is the gravitational constant, $\alpha$ is the (dimensionless) strength of the Yukawa potential relative to Newtonian gravity, and $\lambda$ is its range. Note that in this paper, we focus on composition-independent violations of Newtonian dynamics, meaning that the coupling constant $\alpha$ does not depend on the test mass used for the experiment. Composition-dependent tests exist, which assume that $\alpha$ depends on the nature of the test masses; in practice, they test the weak equivalence principle (see \citet{berge18} and \citet{fayet19} for up-to-date constraints on composition-dependent tests).

\begin{figure}[t]
\includegraphics[width=0.9\textwidth]{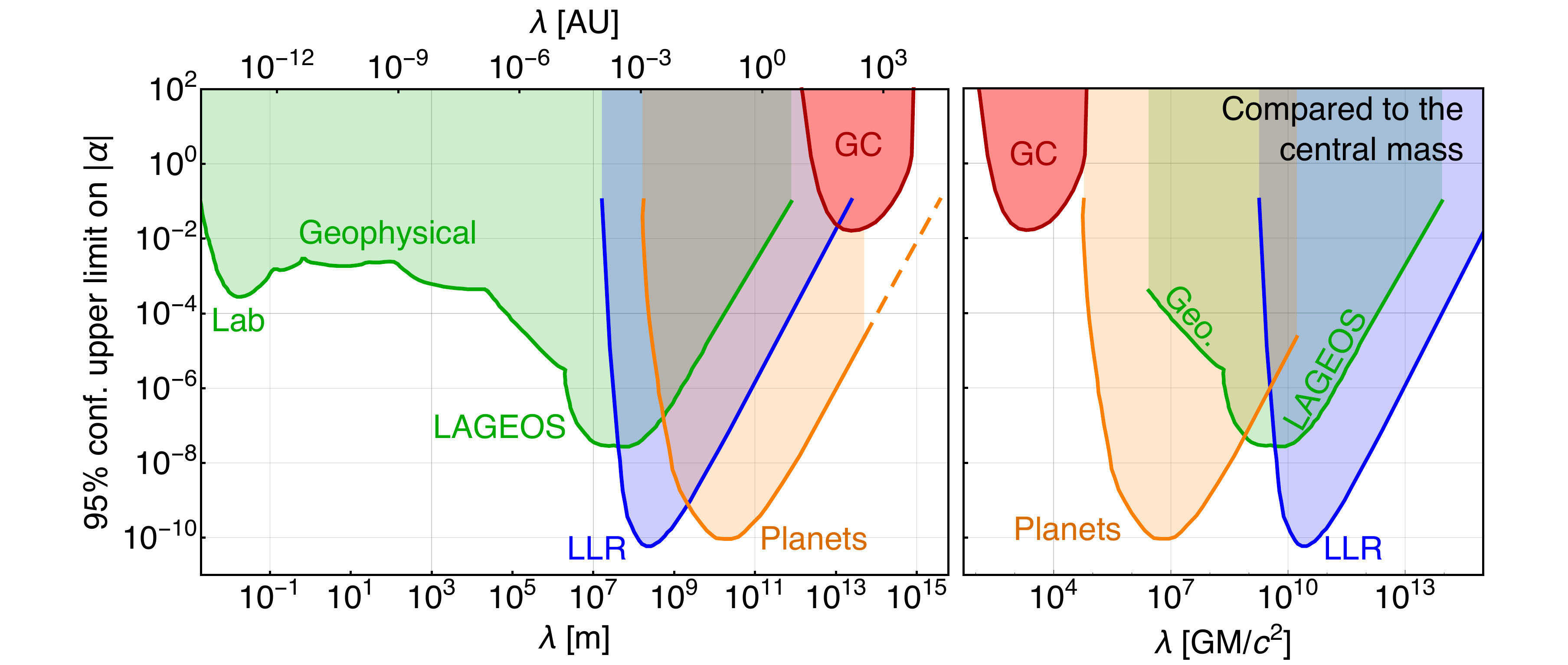}
\caption{\small 95\%-confidence-level constraints on ISL violating Yukawa interactions, from millimeter scales to 1000 AU. Figure from \citet{hees17} and \citet{konopliv11}.}
\label{fig_yukawa_constraints}       
\end{figure}

Fig. \ref{fig_yukawa_constraints} shows the current constraints on ISL-violations from a Yukawa potential, from millimeter scales to 1000 AU. 
A large region of the ($\alpha$, $\lambda$) plane is still allowed by current experiments at submillimeter scales. Significant efforts are done to better constrain this region, most notably with torsion pendulums, micro-cantilever or Casimir-force experiments \citep{adelberger03, adelberger09, murata15}. Although these scales may be probed by bringing laboratory experiments to space (albeit it is not clear whether the performance would be improved --\citet{island}), we do not explore this possibility in this white paper, but only concentrate on ranges larger than planetary scales. Note that geophysical scales can be probed by torsion pendulums using the Earth as the attractor (e.g. \citet{wagner12}), while constraints on earth-orbit scales are provided by a wealth of Earth-orbiting satellite experiments \citep{iorio02, lucchesi11, ciufolini13, delva17}, though care should be taken to properly account for the shape of the Earth \citep{berge18b}.

Larger scales (from a few to hundreds of AU) can only be probed by Solar System experiments, since astrophysical and cosmological observations are by definition sensitive to much larger scales.
The tightest constraints come from measurements by the Lunar Laser Range (LLR --e.g. \citet{murphy13}). The LLR consists in range measurements of the Moon, so that its orbit is extremely finely monitored, with a subcentimetric precision. The most sensitive observable for testing the ISL with LLR is the anomalous precession of the lunar orbit, due to the Earth's quadrupole field, perturbations from other solar system bodies, and relativistic effects. Those perturbations and effects can be parametrized as a Yukawa interaction, thereby providing the constraints shown in Fig. \ref{fig_yukawa_constraints}. 

Larger scales can be probed through planetary ephemeris \citep{fienga10,viswanathan17} and ranging. Ranging is done by bouncing radar signals onto planets, or using microwave signals transmitted by orbiting spacecrafts. As shown in Fig. \ref{fig_yukawa_constraints}, such techniques provide a tight constraint on $\alpha$ \citet{konopliv11}. However, this constraint becomes quickly loose as the scale increases towards the outer solar system scale.

Probing those scales (the largest scales reachable by man-made instruments in a reasonable amount of time) can be done by measuring the orbit of an outbound spacecraft, as it drifts away from the Sun and the planets \citep{jaekel05, hees12}. The most notable existing test in this domain was performed by NASA during the extended Pioneer 10 \& 11 missions. The test resulted in the so-called Pioneer anomaly  \citep{anderson98, anderson02a}, finally accounted for by an anisotropic heat emission from the spacecrafts themselves \citep{rievers11, turyshev11}. 

The effect of non-gravitational forces could be definitely accounted for if, instead of relying on spacecraft and environment models, whose accuracy can always be doubted, we were able to measure them directly. This can be easily done with an onboard accelerometer, as proved by as many missions as GOCE \citep{rummel11}, LISA Pathfinder \citep{armano16} and MICROSCOPE \citep{touboul17}. 
Combining the model-independent measurements of non-gravitational accelerations provided by an on-board accelerometer with radio tracking data, it becomes possible to improve by orders of magnitude the precision of the comparison with theory of the spacecraft gravitational acceleration. 
Such experimental concepts have been proposed e.g. by \citet{christophe09} and \citet{buscaino15}.
This is an aspect of the mission outline that we propose in this white paper.

\subsubsection{Gravitation's low-acceleration frontier}

\begin{figure}[t]
\includegraphics[width=0.9\textwidth]{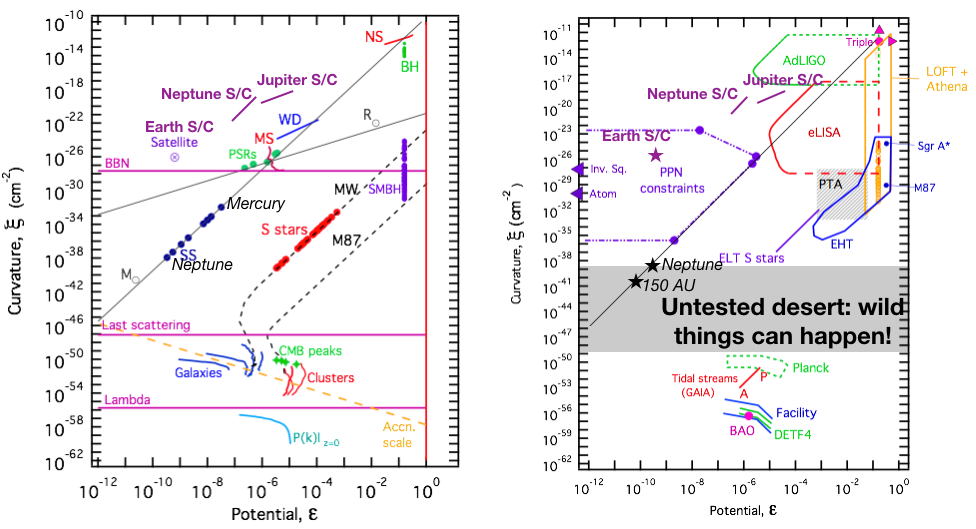}
\caption{\small Gravitational potential--curvature plane. {\it Left}: Astrophysical, cosmological and experimental probes. {\it Right}: Observationally and experimentally tested (currently or planed) regions are shown in color; the shaded region is the so called ``desert", where no experimental is currently available, but where a new intermediate regime of gravitation could be found, bridging between GR at higher curvature and ``dark energy gravitation" at smaller curvature. Exploring this desert should be set as a priority. Figure adapted from \cite{baker15}.}
\label{fig_desert}       
\end{figure}

\citet{baker15} classify probes and experimental/observational tests of gravitation in the potential--curvature plane (left and right panels of Fig. \ref{fig_desert}, respectively). There, the potential $\epsilon$ and curvature $\xi$ are loosely defined as the Newtonian gravitational potential
\begin{equation}
\epsilon \equiv \frac{GM}{rc^2}
\end{equation}
and the Kretschmann scalar
\begin{equation}
\xi \equiv \left( R^{\alpha \beta \gamma \delta} R_{\alpha \beta \gamma \delta}\right)^{1/2} = \sqrt{48} \frac{GM}{r^3c^2}
\end{equation}
created by spherical body (Schwarzschild metric) of mass $M$ at a distance $r$, where $c$ is the speed of light and $R$ is the Ricci curvature tensor.

It is clear that this plane is divided in four main regions: 
\begin{itemize}
\item highest curvatures and potentials correspond to compact objects and can be tested with gravitational waves observatories
\item smaller curvatures correspond to Solar System objects (small potential) and to Galactic center's S-stars (higher potential); the former can be tested with planets ephemeris and man-made spacecrafts (see above), while the latter can be tested with Galactic center observations
\item very small curvatures correspond to cosmological probes (galaxies, large scale structures) and can be tested e.g. with CMB observations or galaxy surveys
\item a desert of probes and (most importantly) of tests lies between the Solar System scale tests and cosmological tests. All kind of speculation can be allowed in this regime, and we can easily imagine that gravitation enjoys a gradual change of regime between compact objects and Solar System scales (``high" curvature, where GR holds) and cosmological scales (``very low" curvature, where GR seems to break), without even needing any of the screening mechanisms mentioned in Sect. \ref{sect_de}.
\end{itemize}

The Solar System tests that we discussed above are shown by the ``Earth S/C" symbols and the filled circles lying on the black slanted line (that stands for the Sun as the source of gravitation). It is clear that in order to barely approach the potential--curvature desert, we must precisely monitor how trans-Neptunian objects (either planetoids or spacecrafts) behave under the influence of the Sun's gravity. Having a test-mass at least 150 AU from the Sun would allow us to actually enter that uncharted desert. 
Additionally, when parametrizing a deviation from Newtonian gravity as a Yukawa potential, we can see from Fig. \ref{fig_yukawa_constraints} that flying a spacecraft at 150 AU from the Sun and farther would test a fifth force of range for which no constraints exist yet.

In Fig. \ref{fig_desert}, we added the regions that could be tested by monitoring the motion of a spacecraft in the Jupiter and Neptune systems. Although such measurements would probe still untested regions of the potential--curvature plane, they remain even farther than tests in the Earth orbit from the potential--curvature desert: albeit interesting in their own, their constraining and exploratory potential on the fundamentals of gravitation is much less than having a test mass 150 AU from the Sun.

\bigskip
We can then state the first goal of our ``dark sector" science case as: 

\noindent {\bf Science goal \#1--} {\it Begin the exploration of the potential--curvature desert with a man-made spacecraft.}  This can be done by directly measuring the gravitational potential with an atomic clock on-board the spacecraft, and by comparing the spacecraft's trajectory with that predicted by GR through the combination of ranging data and the in-situ measurement (and correction) of non-gravitational accelerations with an on-board accelerometer (Sect. \ref{sect_missionconcept}).
Only by entering this potential--curvature desert will we be able to gain new experimental clues about gravity, and perhaps about physics beyond GR. We could in particular enter a new intermediate regime that bridges between GR in the Solar System and ``dark energy gravitation"  at cosmological scales.
Additionally, we would approach MOND's acceleration, thereby allowing us to directly test MOND \citep{banik19}.

\subsection{Dark matter density measurement}

\subsubsection{(Direct) detection techniques and their link to the solar neighborhood}

Although dark matter mostly interacts gravitationally with baryonic matter, it is expected that it leaves tiny signatures observable non-gravitationally. Thus, indirect and direct detections methods have been devised to hunt for it. 

Indirect detection relies on observing electromagnetic signals or neutrinos originating from the annihilation of dark matter particles (see e.g. \citet{slatyer17} for a review). We will not deal with such indirect detection techniques in the remainder of this paper.

Direct detection techniques have been invented to look for different species of dark matter: axions are looked for via their resonant conversion in an external magnetic field using microwave cavity experiments \citep{kuster08}, while WIMPs are expected to occasionally interact with heavy nuclei, thereby creating a detectable nuclear recoil (see \citet{schumann19} for a review of existing experiments); moreover, light dark matter (below the GeV scale) can be searched via scatters off electrons \citep{essig12}, while ultra-light dark matter can cause tiny but apparent oscillations in the fundamental constants that can cause minute variations in the frequency of atomic transitions \citep{arvanitaki15, stadnik15,hees16} and a fifth force (e.g. \citet{berge18}).

\begin{figure}[t]
\includegraphics[width=0.53\textwidth]{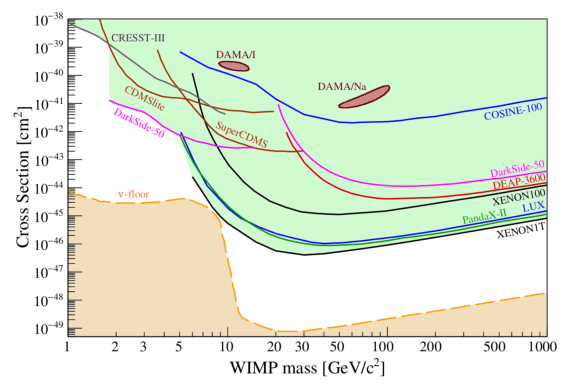}
\caption{\small Current limits on spin-independent WIMP-nucleon scattering experiments \citep{cresstIII19,cdmslite16,darkside18,superCDMS18,dama18,cosine100,deap3600,xenon100,lux17,panda17,xenon100,xenon1t}. The orange shaded region represents the neutrino floor from the irreducible background from coherent neutrino-nucleus scattering.
Figure from \citet{schumann19}.}
\label{fig_wimp}       
\end{figure}

For illustrative purpose, we now restrict our discussion to the spin-independent direct detection of WIMPs, through the elastic scattering off nuclei. 
Fig. \ref{fig_wimp} shows the current limits in the WIMP mass -- WIMP-nucleon cross-section plane from several direct detection experiments.
The differential rate for WIMP scattering is degenerate between the local WIMP density in the galactic halo $\rho_0$ and the WIMP-nucleus differential cross-section ${\rm d}\sigma/{\rm d}E_R$ \citep{baudis12,schumann19}
\begin{equation} \label{eq_dm}
\frac{{\rm d}R}{{\rm d}E_R} = N_N \frac{\rho_0}{m_w} \int_{v_{\rm min}}^{v_{\rm max}} {\rm d}{\bf v} f({\bf v}) v \frac{{\rm d}\sigma}{{\rm d}E_R},
\end{equation}
where $N_N$ is the number of target nuclei, $m_w$ is the WIMP mass, $E_R$ is the energy transferred to the recoiling nucleus, and ${\bf v}$ and $f({\bf v})$ are the WIMP velocity and velocity distribution in the Earth frame. The lowest allowed velocity $v_{\rm min}$ is that needed for a WIMP to induce a detectable nuclear recoil, while the maximum one $v_{\rm max}=533^{+54}_{-41}$km/s corresponds to the escape velocity \citep{piffl14}, meaning that WIMPs with $v>v_{\rm max}$ are not bound to the Galaxy.

The WIMP velocity distribution $f({\bf v})$ must be estimated with numerical simulations \citep{vogelsberger09, bozorgnia16, sloane16, helmi18}, though new observations could constrain $f({\bf v})$ \citep{ruchti15}. Since it cannot be determined through the dynamics of a spacecraft, we will ignore it in the remainder of this paper, but instead focus on how and why we can measure the local dark matter density $\rho_0$. We should first note that current estimates mostly agree in the range $\rho_0=(0.4-1.5)$ GeV/c$^2$/cm$^3$ \citep{sivertsson18, buch19}.

It is clear from the degeneracy in Eq. (\ref{eq_dm}) that, were $\rho_0$ revised, upper limits of Fig. \ref{fig_wimp} would be affected \citep{green17}. It is therefore particularly important to have an estimate as precise as possible of $\rho_0$. Two traditional ways to measure it have been used for decades by astronomers. We propose a novel, in-situ way, in Sect. \ref{ssect_sclocal}. But first, we summarize those two traditional observational ways (see e.g. \citet{read14} and references therein).

The local dark matter density is an average of the Galactic dark matter halo density in a small region about the Solar System (at 8 kpc from the center of the Galaxy), extrapolated to the laboratory.
The first method to estimate it then relies on the measurement of the rotation curve of the Galaxy and on a model of the dark matter halo. The halo is usually taken to be an isothermal sphere with density $\rho(r) \propto r^{-2}$, so that we recover the flat rotation curve, but can be parameterized as a better model such as a Navarro-Frenk-White (NFW) halo \citep{nfw97}. A bad model choice is obviously a source of systematic uncertainties. 
The second method does not rely on a halo model, but uses the vertical motion of stars close to the Sun. The main systematics come from the distance and velocity measurement of those stars (so that K-stars are preferentially chosen), though robust techniques have been developed \citep{garbari12}.
Combining the two methods allows astronomers to assess whether the Galactic halo is prolate or oblate, and to look for a dark disk.

Since the measured $\rho_0$ must be extrapolated to the position of the Earth, the local homogeneity of the dark matter halo is an important point. In particular, any clump or stream of dark matter, as well as dark matter trapped by the Sun's potential \citep{peter09} may bias the extrapolation. Although dark matter simulations' resolution is no better than a dozen parsecs, we have good reasons to assume that the possible effects are negligible, and to consider the local dark matter halo as homogeneous \citep{read14}. The presence of a dark disk on top of the halo should not cause a problem either, as it was shown that if such a disk exists, it is thick enough not to influence the homogeneity of the close-by dark matter \citep{read14}.

We shall close this presentation with a major caveat \citep{read14, green17, evans19}: although measurements of $\rho_0$ relying on dark-matter-only simulation are very robust, adding baryons to the simulations significantly complicates things. However, baryons are clear components of our neighborhood, and therefore must be accounted for.

\subsubsection{In-situ measurement of the local dark matter density} \label{ssect_sclocal}

Tracking a spacecraft far enough from the Sun, we can in principle estimate the contribution of the Galactic dark matter to its dynamics. We then use the same techniques as those presented above, with the stellar tracers replaced by the spacecraft.

The discussion above also highlighted two possible (linked) caveats: (i) speculating on the homogeneity of the local dark matter distribution to estimate its density in Earth laboratories from astrophysics estimates averaged on the Solar System neighborhood and (ii) adding baryons to dark-matter-only simulations complicates predictions about the Galactic dark matter halo. 
These caveats highlight the difficulty to calibrate models without an in-situ measurement of physical quantities. This is especially true in our Galactic neighborhood, since it is dominated by baryons, so that hunting for dark matter is particularly difficult.

\begin{figure}[t]
\includegraphics[width=0.53\textwidth]{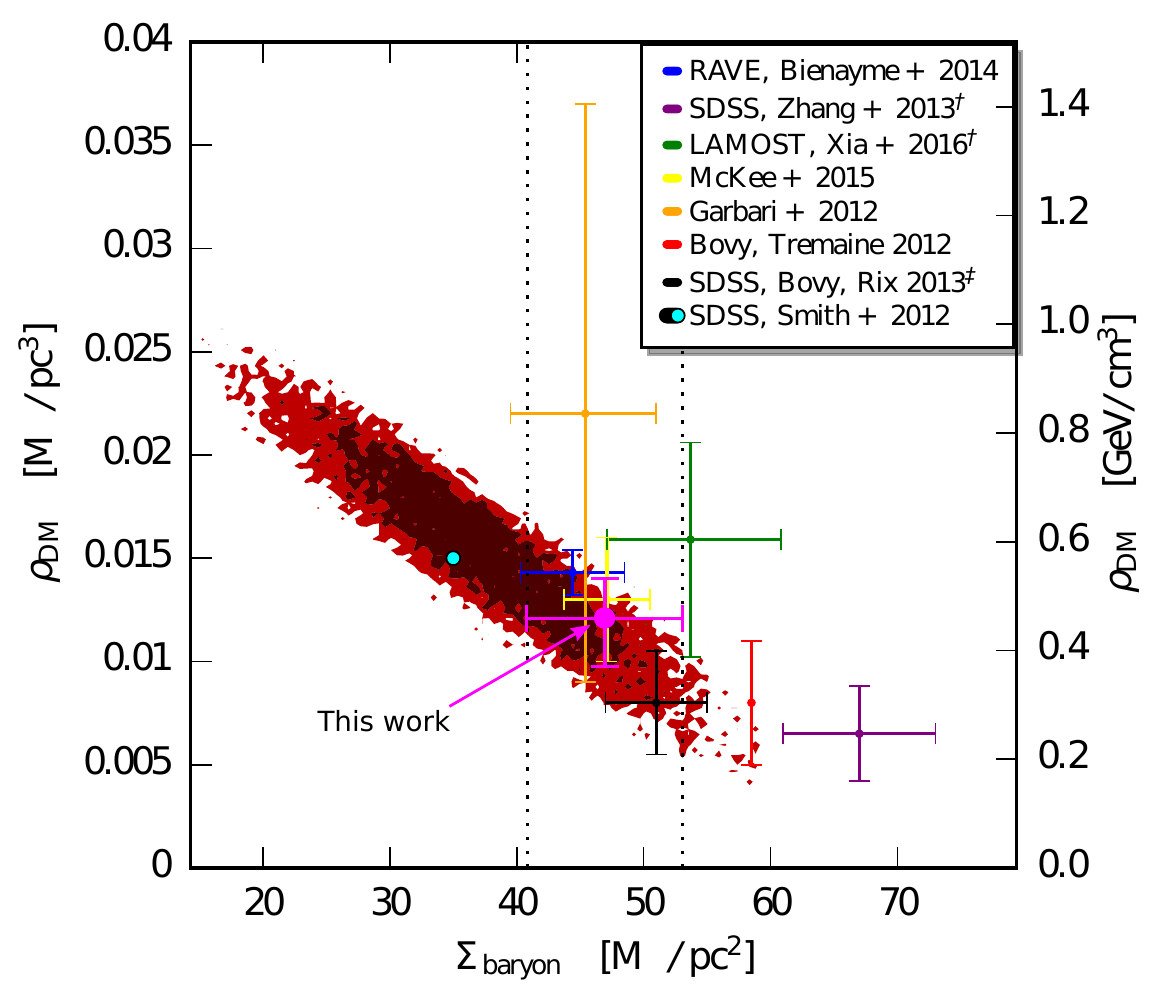}
\caption{\small Degeneracy between the local dark matter density and the baryonic contribution to the local dynamical mass. Figure from \citet{sivertsson18}, with \citet{bienayme14, zhang13, xia16, mckee15, garbari12, bovy12}.}
\label{fig_read}       
\end{figure}

The interplay between dark matter and baryons is most easily seen by the uncertainty on $\rho_0$ steming from our imperfect knowledge of the baryonic  contribution to the local dynamical mass, as shown with the baryonic surface density $\Sigma_b$ in Fig. \ref{fig_read} \citep{sivertsson18}.

Better understanding the respective contributions and distributions of dark matter and baryons through experimental in-situ measurements is not only important for direct detection experiments (as it allows us to better estimate $\rho_0$) but also to improve our knowledge of our local environment. Although dark-matter-only simulations expect a smooth dark matter distribution, we could be surprised to discover small clumps, streams and passing clouds of dark matter: indeed, as aforementioned, ultra-light dark matter (clumped e.g. in the form of topological defects or axion stars) can cause minute variations in the frequency of atomic transitions; an atomic clock going through such a cloud would be temporarily desynchronized compared to Earth-bound clocks \citep{derevianko14, wcislo16, roberts17}. Only a direct assessment of the gravitational environment at very small scales will allow us to close this question: as dark matter can affect the rate of a clock, it can be done with a spacecraft carrying a clock, as shown in Sect. \ref{sect_missionconcept}; if the clump is sufficiently massive, its crossing can be determined by the precise monitoring of the spacecraft's trajectory.

Regarding baryonic matter, an accelerometer on-board a spacecraft measures all non-gravita\-tio\-nal forces applied to the satellite (gas and dust friction, solar radiation pressure, outgasing...). It is then an invaluable tool to directly probe the baryons distribution along its trajectory. In other words, it provides an in-situ experimental measurement of the local contribution of dust and gas in $\Sigma_b$. 

Such measurements will also be valuable for cosmology, as they could be used to calibrate galaxy evolution simulations, and to better understand the systematics that still plague weak lensing surveys \citep{rudd08,semboloni11}.

\bigskip
We can then state the second goal of our ``dark sector" science case as: 

\noindent {\bf Science goal \#2--} {\it Improve our knowledge of the local dark matter and baryon densities}
It can be fulfilled by monitoring the dynamics of a spacecraft in the Solar System neighborhood, the spacecraft carrying a clock and an accelerometer. The clock will be sensitive to local dark matter inhomogeneities. The combination of ranging and accelerometric data will also be sensitive to local gravitational disturbances, such as those that could be created by a massive enough clump. Finally, accelerometric data will be sensitive to the friction of any baryonic matter (dust and gas) on the spacecraft, allowing for a direct measurement of the baryonic matter density along the spacecraft trajectory. This will allow us to perform the first truly local measurement of the dark matter halo density $\rho_0$ and to improve the characterization of the dark matter constraints from direct detection experiments.

\section{A very deep space mission concept} \label{sect_missionconcept}

In this section, we present the concept of a mission that could fulfill the science goals presented above.

\subsection{Overview and requirements}

%
%

\subsubsection{Science goal \#1: low-acceleration gravitation}

This goal can be fulfilled by directly measuring the gravitational potential and by looking for a deviation from the ISL.

\paragraph{Gravitational potential}

The universal redshift of clocks when subjected to a gravitational potential is one of the key predictions of General Relativity (GR) and, more generally, of all metric theories of gravitation. It represents an aspect of the Einstein Equivalence Principle (EEP) often referred to as Local Position Invariance (LPI) \citep{will14}, and it makes clocks direct probes of the gravitational potential (and thus to the metric tensor $g_{00}$ component).
In GR, the frequency difference of two ideal clocks is proportional to \citep{reynaud09}
\begin{equation} \label{eq_lli}
\frac{{\rm d} s_g}{c {\rm d}t} - \frac{{\rm d} s_s}{c {\rm d}t} \approx \frac{w_g - w_s}{c^2} + \frac{v_g^2 - v_s^2}{2c^2},
\end{equation}
where $w$ is the Newtonian potential and $v$ the coordinate velocity, with $g$ ad $s$ standing for the ground and the in-space clocks.
In theories different from GR this relation is modified, leading to different time and space dependence of the frequency difference. This can be tested by comparing two clocks at distant locations (different values of $w$ and $v$) via exchange of an electromagnetic signal. 

An outbound Solar System trajectory (large potential difference) and low uncertainty on the observable (\ref{eq_lli}) allows for a relative uncertainty on the redshift determination given by the clock bias divided by the maximum value of $\Delta w/c^2$.
Assuming that the Earth station motion and its local gravitational potential can be known and corrected to uncertainty levels below $10^{-17}$ in relative frequency (10 cm on geocentric distance), which, although challenging, are within present capabilities, then for a onboard clock similar to ACES' PHARAO (see below), with a $10^{-17}$ bias \citep{reynaud09}, at a distance of 150 AU this corresponds to a test with a relative uncertainty of $10^{-9}$, an improvement by almost four orders of magnitude on the uncertainty obtained by the currently most sensitive experiments \citep{delva18, hermann18}.

We can thus take as a requirement that the clock's bias is less than $10^{-17}$.

\paragraph{Inverse Square Law violation}

A definitive deviation from the ISL can be detected as a deviation from the trajectory predicted by GR when taking into account the gravity of the Solar System's bodies. What is needed is an accurate orbit restitution, making sure that the spacecraft follows a geodesics. The former can be done through orbit tracking with Radio-Science, while the latter can be ensured with a drag-free spacecraft, whose trajectory is forced to be a geodesics by actively canceling non-gravitational forces; alternatively, we can measure the non-gravitational forces with a DC accelerometer, and correct for them when estimating the orbit, therefore not needing a drag-free spacecraft. Although a model of non-gravitational forces is commonly used to correct for them, we argue that no model is worth an empirical measurement, and hence that an accelerometer (or drag-free spacecraft) is needed to definitely confirm any measured deviation from the ISL.

The required accuracy of non-gravitational forces is driven by the accuracy on the orbit estimation and depends on the time of integration for the orbit restitution: the deviations from a GR should be searched for in as short as possible time segments to minimize the effect of the drift. As shown by \citet{hees12}, a one-meter deviation from a Keplerian orbit can be detected in a few days, requiring a precision on non-gravitational forces of the order of $10^{-12}$ m/s$^2$. Getting down to such a precision would significantly improve the current constraints given by the Pioneer probes, which assumed a bias in acceleration of $10^{-10}$ m/s$^2$.

The requirements for the large scale test of the ISL can then be summarized as \citep{hees12, buscaino15}:

\begin{itemize}
\item measure the orbit of the spacecraft with a precision better than 1m, all along the mission, from the Earth to hundreds of AU
\item measure non-gravitational accelerations at the order of $10^{-12}$ m/s$^2$
\end{itemize}


\subsubsection{Science goal \#2: dark matter: search and local density measurement}

The same experimental apparatus as used for Science goal \#1 is able to fulfill this goal. Although precise requirements must still be computed, we can expect that the clock will detect inhomogeneities in the dark matter distribution. While the accelerometer will be perfectly suited to measure the baryon distribution through the friction applied to the spacecraft, combining its measurements with ranging data should enable us to detect massive enough inhomogeneities in the gravitational field, possibly originating from dark matter clumps or streams.

\subsection{Mission profile}

The main factor limiting the viability of very deep space missions with conventional propulsion systems and space navigation techniques is time. It took between 20 and 25 years to the Voyager and Pioneer probes to reach 50 AU; New Horizons, albeit faster, took 15 years. Drawing on these numbers, reaching 100 AU and more is prohibitively long.

Therefore, if we want to reach gravitation's low-acceleration regime, we must envision new propulsion methods.
Several innovative concepts have been put forth to send interstellar nanoprobes. For instance, the Breakthrough Starshot project aims to send a nanoprobe (of about 10 g) to Proxima Centauri. In order to reach this star (at 4.5 light-years), the probe will be given a strong impulse with a laser (shot from the ground) hitting a solar sail. The designers of the project expect the probe to reach 15\% of the speed of light.

We can straightforwardly build on this concept to design  a very deep mission aiming to test the dark sector in the outskirts of the Solar System. Let $m$ and $M$ be the mass of Breakthrough Starshot's probe and of our spacecraft, respectively. Then, assuming the velocity of the Breakthrough Starshot probe is $0.15c$ and that the same amount of energy is transferred to our spacecraft (with a solar sail) by the ground laser, then its velocity is simply
\begin{equation}
v = 0.15 \sqrt{\frac{m}{M}}c.
\end{equation}
Assuming $m=10$ g, then a spacecraft of mass between 100 kg and 1000 kg should be able to reach 150 AU in 2 years to 6 years.

The mission profile is then straightforward. First, launch the spacecraft in Low Earth Orbit. As for MICROSCOPE, it could be passenger of a Soyuz-like launcher. Second, accelerate it with Breakthrough Starshot's laser to put it into a high-velocity Solar System outbound trajectory. A gravitational boost with Jupiter or Saturn could be envisioned to gain more speed in the first stage of its cruise.

This mission concept crucially depends (and speculates) on the availability of the Breakthrough Starshot (or similar) laser propulsion system. If such a system exists in the 2040s, can be used intensively (thereby, of a low usage cost), then our concept could fit into a M-class mission.

We detail the possible payload and platform below. We then conclude the section with the discussion of possible secondary science goals.

\subsection{Payload}

\subsubsection{Clock}

As shown above, a clock is directly sensitive to the gravitational potential. We thus propose to embark an atomic clock, the design of which can be based on the PHARAO clock, set to fly as part of the ACES experiment on-board the International Space Station in 2020 \citep{reynaud09}.

In microgravity, the linewidth of the atomic resonance of the clock will be tuned by two orders of magnitude, down to sub-Hertz values (from 11 Hz to 110 mHz), 5 times narrower than in Earth based atomic fountains. After clock optimization, performances in the $10^{-16}$ range are expected both for frequency instability and inaccuracy.

Developed by CNES, the cold atom clock PHARAO combines laser cooling techniques and microgravity conditions to significantly increase the interaction time and consequently reduce the linewidth of the clock transition. Improved stability and better control of systematic effects will be demonstrated in the space environment.
PHARAO can reach a fractional frequency instability of $10^{-13}  \tau^{-1/2}$, where $\tau$ is the integration time expressed in seconds, and an inaccuracy of a few parts in $10^{16}$.

Frequency transfer (via microwave link) with time deviation better than 0.3 ps at 300 s, 7 ps at 1 day, and 23 ps at 10 days of integration time will be demonstrated by ACES. 


\subsubsection{Electrostatic absolute accelerometer}

Correcting for non-gravitational accelerations can be done with an unbiased (DC) accelerometer. One is currently under development at ONERA: the Gravity Advanced Package (GAP) is composed of an electrostatic accelerometer (MicroSTAR), based on ONERA's expertise in the field of accelerometry and gravimetry (CHAMP, GRACE, GOCE and MICROSCOPE missions), and a bias calibration system \citep{lenoir13}. Ready-to-fly technology is used with original improvements aimed at reducing power consumption, size and weight. The bias calibration system consists in a flip mechanism which allows for a 180$^{\rm o}$ rotation of the accelerometer to be carried out at regularly spaced times. The flip allows the calibration of the instrument bias along 2 directions, by comparing the acceleration measurement in the two positions.

\begin{figure}[t]
\includegraphics[width=0.25\textwidth]{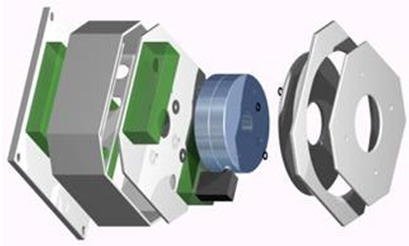}
\unskip\ \vrule\
\reflectbox{\rotatebox[origin=c]{360}{\includegraphics[width=0.4\textwidth]{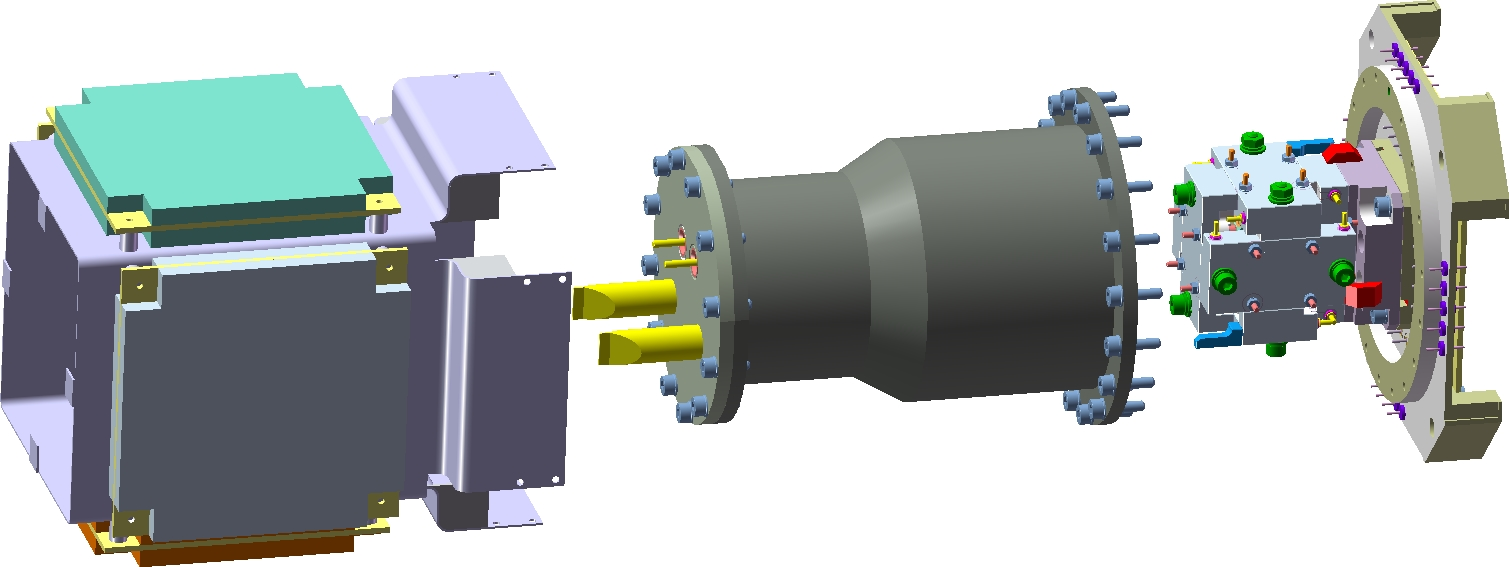}}}
\unskip\ \vrule\
\includegraphics[width=0.2\textwidth]{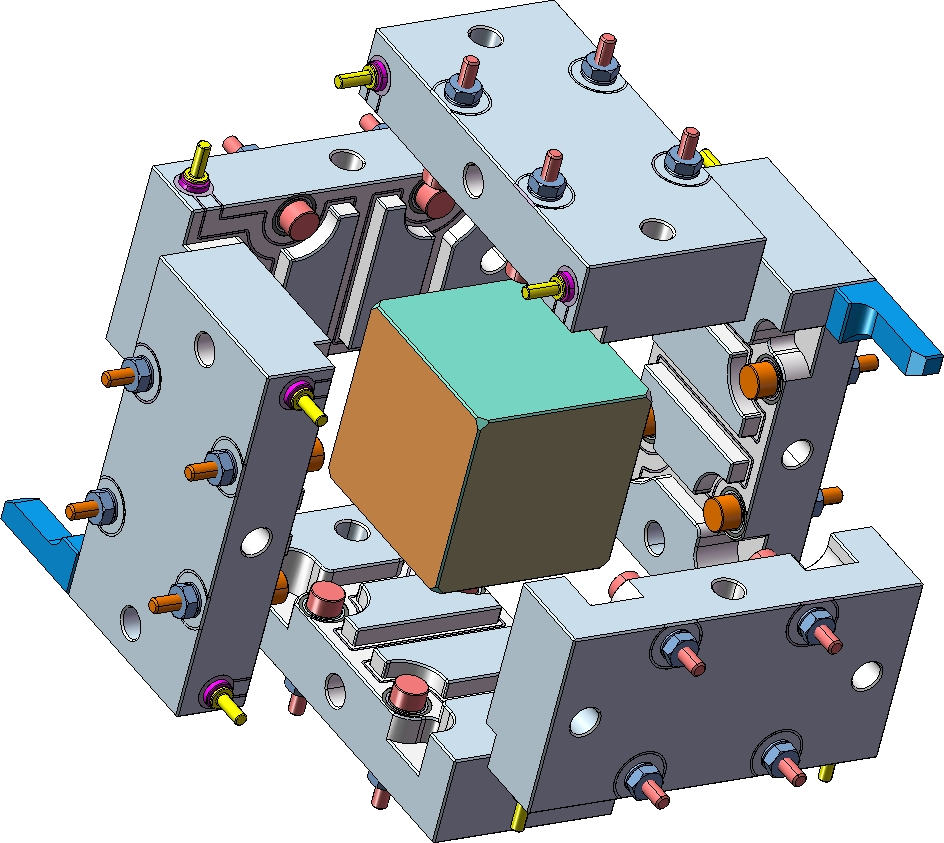}
\caption{\small GAP DC accelerometer. {\it Left}: Bias Rejection System. {\it Center}: Exploded view of MicroSTAR accelerometer. {\it Right}: Exploded view of MicroSTAR's test mass in its cage.}
\label{fig_gap}       
\end{figure}

The three axes electrostatic accelerometers developed at ONERA are based on the electrostatic levitation of the instrument inertial mass with almost no mechanical contact with the instrument frame. The test-mass is then controlled by electrostatic forces and torques generated by six servo loops applying well measured equal voltages on symmetric electrodes. Measurements of the electrostatic forces and torques provide the six outputs of the accelerometer. The mechanical core of the accelerometer is fixed on a sole plate and enclosed in a hermetic housing in order to maintain a good vacuum around the proof-mass (Fig. \ref{fig_gap}, middle and right panels). The electronic boards are implemented around the housing. The control of the proof-mass is performed by low consumption analog functions.

The Bias Rejection System is equipped with a rotating actuator and a high-resolution angle encoder working in closed loop operation (Fig. \ref{fig_gap}, left). 
The actuator for the MicroSTAR SU rotation is a stepper motor with worm gear.
The electronic boards for driving the stepper motor and controlling the closed servo loop are located inside the housing. The electrical connection between Sensor Unit and Interface Control Unit is given by thin flexible shielded wires designed for the corresponding environment and durability requirements. Sliding contacts are not necessary because the maximum turning angle is 270$^{\rm o}$ to one direction. A second small actuator (linear servo motor) is foreseen to block the moving part of the rotating actuator during transport and launch.

\begin{figure}[t]
\includegraphics[width=0.53\textwidth]{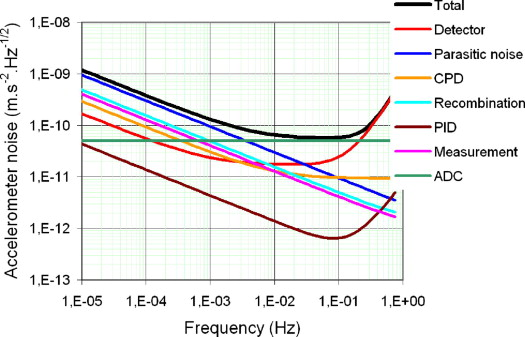}
\includegraphics[width=0.4\textwidth]{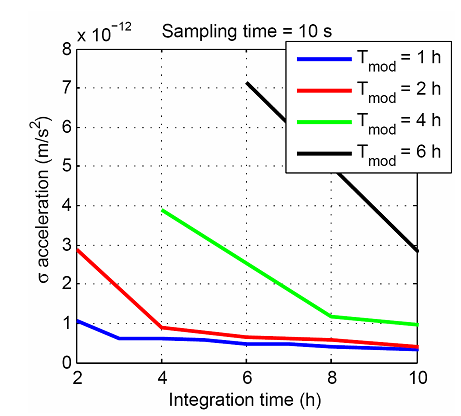}
\caption{\small Current GAP performance. {\it Left}: MicroSTAR's noise. {\it Right}: Bias rejection system's resolution.}
\label{fig_gap_perfo}       
\end{figure}

The global performance of the acceleration measurement is allocated towards the different contributors that appear in the expression of the measurement equation. Eight posts are defined with respect to the different contributors: at first order (assuming that the non-gravitational acceleration $a_{\rm NG}$ is the dominant term in a perfect measurement), the measured acceleration is:
\begin{multline}
a_{\rm meas} = a_{\rm NG} + n + b + a_{\rm parasitic, acc} + [K] a_{\rm NG} + [K2] a^2_{\rm NG} + [R+S] a_{\rm NG} + ([\dot{\Omega}] + [\Omega^2]) r \\
+ a_{\rm parasitic, SC} - [U]r + 2 [\Omega]\dot{r} + \ddot{r},
\end{multline}
where $n$ is the instrumental noise, $b$ MicroSTAR's bias, $a_{\rm parasitic, acc}$ some parasitic acceleration acting as a bias, $[K]$ the scale factor, $[K2]$ the quadratic factor, and $[R+S]$ accounts for rotations and misalignments, $[\dot{\Omega}] + [\Omega^2]$ for the inertial acceleration, $a_{\rm parasitic, SC}$ for the spacecraft self-gravity, and $(- [U]r + 2 [\Omega]\dot{r} + \ddot{r})$ describes gravity gradient and deformation.

The left panel of Fig. \ref{fig_gap_perfo} shows the current noise of the MicroSTAR accelerometer, with different contributors: a level of 10$^{-9}$ m/s$^2$/Hz$^{1/2}$ is obtained over [10$^{-5}$ - 1] Hz, with a measurement range of $1.8 \times 10^{-4}$ m/s$^2$. The bias modulation signal, consisting in regular 180$^{\rm o}$ flips of the accelerometer, allows us to band-pass filter the accelerometer noise around the modulation frequency. The right panel of Fig. \ref{fig_gap_perfo} shows, for different periods of modulation and calibration, that after rejecting the bias, GAP is able to measure absolute accelerations down to $10^{-12}$ m/s$^2$.

\subsubsection{Tracking, clock synchronization and communication}

Although the precision attainable on the orbit determination at distances larger than 150 AU is uncertain, the spacecraft's orbit can be determined through ranging, Doppler tracking and Very Long Baseline Interferometry (VLBI).

Ranging should rely on the Deep Space Network (DSN), through the measurement of the time an upward signal takes to be retransmitted by the spacecraft. The power necessary to run the transponder onboard the spacecraft, as well as the antenna gain, are at the moment not precisely known. The return on experience of the Voyager and New Horizon probes will be beneficial to quantify them.

A microwave link will enable Doppler tracking of the spacecraft and the synchronization of the onboard clock with on-ground clocks.

Orbit determination should be performed regularly, however we do not expect that a very large duty cycle will be necessary. Therefore, this mission should not need a high access to DSN facilities.

\subsection{Platform}

The spacecraft should provide the lowest and most axisymmetrical gravitational field, and make coincide as much as possible the dry mass center of gravity, the propellant center of gravity, the radiation pressure force line and the accelerometer.
Furthermore, the satellite should be as rigid as possible to minimize the impact of vibrations and glitches due to mechanical constraints, which could contaminate the science measurements. A platform similar to that of MICROSCOPE could be envisaged. Thence, its weight can be expected about 500 kg.

Radioisotope Thermoelectric Generators (RTG) are required to operate a spacecraft beyond Jupiter. 
Furthermore, it is known that RTG can create non-gravitational forces through the radiations they emit (they are the best candidates to explain the Pioneer anomaly --\citet{rievers11, turyshev11}). The onboard accelerometer will solve this problem.

A propulsion module will be needed only if deep-space maneuvers are required. This will depend on the orbit definition, and may largely affect the design of the spacecraft. To minimize the self-gravity nuisance from a propulsion module in the spacecraft, we could envision using an external propulsion module to set the spacecraft on its final orbit, that would be eventually released once the orbit is reached.


The attitude of the spacecraft may be passively controlled, at least in part, by spinning it slowly about its principle axis of symmetry; counter-rotating the payload would allow us to perform tests in a non-rotating frame.
The attitude can be estimated with stellar sensors combined with the onboard accelerometers.

\subsection{Secondary objectives}

A space mission as that presented above allows for several secondary objectives:

%
%

\paragraph{Science goal \#3-1}
{\it Measurement of Post-Newtonian parameters}

\noindent The Parametrized Post Newtonian (PPN) formalism relies on linearizing Einstein's equations in the weak field and low velocity regime \citep{will14}. A spacecraft leaving the Solar System is in this exact regime, such that the mission concept described above can provide new constraints of those parameters.
For instance, the Eddington parameter $\gamma$ (which measures how much space curvature is produced by unit rest mass) can be estimated during any solar conjunction during the spacecraft's cruise  to repeat the Cassini relativity experiment \citep{bertotti03}.

\paragraph{Science goal \#3-2}
{\it Variation of fundamental constants}

\noindent All deviations from GR mentioned in this white paper can be attributed to the variation of fundamental constants \citep{uzan03}. The space mission concept that we depicted can thus be seen as a direct probe of the evolution of those constants.

\paragraph{Science goal \#3-3}
{\it Kuiper belt exploration}

\noindent Since their first detection \citep{jewitt93}, Kuiper Belt Objects (KBOs) have revolutionized our understanding of the history of the Solar System. Thousands of them have since been discovered, that exhibit remarkable diversity in their properties (see e.g. \citet{brown12} for a review about KBOs). 
A spacecraft passing through the Kuiper belt is affected by its gravitational potential, its shape, and the possible dust distributed in the area. Combined to giant telescopes expected to be online in a few decades, a clock and an accelerometer on-board the spacecraft are the perfect tools to measure them --see e.g. the SAGAS science case \citep{wolf09}.

\paragraph{Science goal \#3-4}
{\it Heliosphere, heliopause and heliosheath}

\noindent Thanks to its onboard accelerometer, the outbound spacecraft presented above will be able not only to monitor the variation of solar radiation pressure as it drifts from the center of the Solar System, but also to measure all non-gravitational forces (solar and interstellar winds) as it crosses the heliopause. It will then largely complement the data already provided by the Voyager missions as they left the Solar System a few years ago.
NASA's Senior Review 2008 of the Mission Operations and Data Analysis Program for the Heliophysics Operating Missions\footnote{http://www.igpp.ucla.edu/public/THEMIS/SCI/Pubs/Proposals and Reports/Senior Review 2008 Report Final.pdf} highlighted the importance of continuing Voyager data to gain in-situ knowledge of the heliosheath.

\paragraph{Science goal \#3-5}
{\it Measurement of gravitational waves background}

\noindent \cite{reynaud08} proposed a new way to set bounds on gravitational waves backgrounds created by astrophysical and cosmological sources by performing clock comparisons between a ground clock and a remote spacecraft equipped with an ultra-stable clock, rather than only ranging to an onboard transponder. Their investigation can be used in a mission concept as that described above.

\section{Scientific landscape in the 2030s} \label{sect_landscape}

In this section, we attempt a short anticipation of the scientific landscape in the 2030s to 2040s. We can identify five possible advances in the fields of interest of this white paper.

\begin{itemize}
\item Test of GR in the strong field regime with LISA: the opening of the gravitational universe with LISA (launched in the early 2030s) will allow for significantly improved tests of GR with compact objects. However, we should not expect that these observations explain dark energy nor dark matter.
\item Improved constraints on dark energy and dark matter: full-sky surveys like LSST, Euclid and WFIRST will be completed. We can expect them to significantly tighten the constraints on the dark energy equation of state and on the distribution of dark matter at cosmological scale. Will they strengthen the GR+$\Lambda$CDM model or find new physics (perhaps in the line of the current tension between local and early-Universe measurements of the Hubble constant --e.g. \citet{riess19}) is a difficult question to answer. In any case, they will remain blind to the gravitational potential--curvature desert mentioned above.
\item Fifth force laboratory and ground tests will have improved and probably excluded more of the small-range interaction parameter space. However, by definition, they will remain blind to those deviations from GR that could be measured at hundreds of AU from the Sun.
\item The dark matter parameter space will be better constrained by direct detection experiments and LHC searches. As for WIMPs, it is possible that the mass -- cross section plane of Fig. \ref{fig_wimp} is completely excluded, down to the neutrino floor, by 2030, or that WIMPs have been detected. Nevertheless, the degeneracy between the local dark matter density and the WIMP cross-section will remain. We do not see any plan to improve it without an in-situ measurement performed by a spacecraft in the neighborhood of the Solar System. Similar arguments can apply to other dark matter candidates (even those that may be invented in the next decade).
\item Giant (ground-based) telescopes and the JWST will be online: they will bring new discoveries of KBOs, that could be complemented by flying through the Kuiper belt.
\end{itemize}

None of those potential advances should close the science goals highlighted in this white paper. It then seems that flying a spacecraft hundreds of AU from the Sun and entering the low-acceleration regime of gravitation is the only way to gain fast insight into the dark sector, even as late as 2040 or 2050.

\section{Technological challenges} \label{sect_challenges}

The payload will be inherited from already flown missions. The accelerometer will be derived from those onboard Earth gravitational field observation missions (GOCE, GRACE, GRACE-Follow) or fundamental physics (LISA Pathfinder). Nevertheless, it should be miniaturized to minimize the spacecraft's total weight. The clock can be derived from the ACES/PHARAO clock, that should fly in the International Space Station in 2020.

The platform should not present any technological challenge, apart from providing a gravitational background as quiet as possible.

We identify a handful of potentially major technological challenges:
\begin{itemize}
\item {\it propulsion}: sending a spacecraft at 150 AU in a relatively short amount of time cannot rely on traditional chemical propulsion and space navigation techniques, so that new innovative propulsion systems must be proposed and developed. Speculating on the development of interstellar (nano)probes accelerated with a ground-based high-power laser, we could bet on the availability of such a laser to sufficiently accelerate a few-hundreds kilograms spacecraft.
\item {\it communication and tracking}: although precise requirements on the orbit determination at distances larger than 150 AU must still be computed, it is unclear whether the current ranging technology will be precise enough. The power of the transponder and the antenna gain will have to be maximized.
\item {\it power}: although RTG are technologically ready for a spacecraft on a fast outbound Solar System orbit, an uncertainty remains about the policies regulating the launch of spacecrafts powered by RTG in a few decades (ESA already banned this possibility).
\item a {\it shield} will be required to send an atomic clock in the outer space. This has never been done. For comparison, ACES will fly in Low Earth Orbit, where such shielding is not necessary.
\end{itemize}

\section{Conclusion} \label{sect_conclusion}

Speculating on the development and availability of new innovative propulsion techniques in the 2040s, we presented open science questions related to the characterization of the dark sector. Flying a spacecraft out of the Solar System would allow us to enter the uncharted gravitation's low-acceleration regime, where we may witness deviations from GR, hinting toward a new theory in which gravitation would depend on the spacetime curvature, to evolve from the well-tested Solar System regime of GR to the puzzling behavior we observe at cosmological scales. Deviations from GR could be efficiently looked for by finely tracking the trajectory of the spacecraft carrying an atomic clock and an accelerometer. The clock directly probes the gravitational potential, while the accelerometer measures all non-gravitational accelerations applied to the spacecraft, allowing us to subtract them and directly compare its trajectory to that predicted by GR.

Moreover, such a mission will have the potential to probe our gravitational environment, i.e. to disentangle the contributions of baryons and dark matter. As a consequence, it will be able to measure in-situ the local density of dark matter, which directly affects the constraints on dark matter's characteristics inferred from on-ground direct detection experiments. It will also allow for the estimation of the homogeneity of the dark matter distribution in the Solar System neighborhood, thereby allowing for a better calibration of cosmological simulations and impacting cosmological surveys still affected by the poorly understood coupling between baryonic and dark matters. Finally, a flying atomic clock could detect clumps of ultralight dark matter via its effects on the frequency on atomic transitions.

Beyond the secondary science goals that we identified (measurement of PPN parameters and of the constancy of fundamental constants, exploration of the Kuiper belt and of the heliosheath, as well as measurement of the gravitational waves stochastic background), such a mission will bring new insights about the low-acceleration regime of gravitation. As stellar systems form in this regime, it is of primary importance to better understand it, at a time where exoplanets are routinely found and provide both new clues and new puzzles about the formation of planetary systems.

We shall conclude this white paper by emphasizing that despite a wealth of new experiments getting online in the near future, that will bring new knowledge about the dark sector, it is very unlikely that the science questions that we presented will be closed. More importantly, it is likely that it will be even more urgent than currently to answer them. Tracking a spacecraft carrying a clock and an accelerometer as it leaves the Solar System may well be the easiest and fastest way to directly probe our dark environment.

\newpage
\bibliography{wp}
\bibliographystyle{aa}

\end{document}